

\def\IR{{\hbox{{\rm I}\kern-.2em\hbox{\rm R}}}}
\def\IB{{\hbox{{\rm I}\kern-.2em\hbox{\rm B}}}}
\def\IC{{\ \hbox{{\rm I}\kern-.6em\hbox{\bf C}}}}

\def\IZ{{\hbox{{\rm Z}\kern-.4em\hbox{\rm Z}}}}

\def\underarrow#1{\vbox{\ialign{##\crcr$\hfil\displaystyle
{#1}\hfil$\crcr\noalign{\kern1pt
\nointerlineskip}$\longrightarrow$\crcr}}}
%

\input phyzzx
\let\a=\alpha   \let\d=\delta \let\e=\epsilon
  \let\q=\theta  \let\k=\kappa
\let\l=\lambda \let\m=\mu \let\n=\nu   \let\r=\rho
\let\s=\sigma     
\let\w=\omega         \let\L=\Lambda
 \let\P=\Pi   \let\F=\Phi

\let\pa=\partial

\def\smIR{\hbox{{\footnotesize\rm I}\kern-.2em\hbox{\footnotesize\rm R}}}
\def\IB{\hbox{{\rm I}\kern-.2em\hbox{\rm B}}}

\def\IC{\ \hbox{{\rm I}\kern-.6em\hbox{\bf C}}}

\def\ID{\hbox{{\rm I}\kern-.2em\hbox{\rm D}}}

\def\IE{\hbox{{\rm I}\kern-.2em\hbox{\rm E}}}

\def\IF{\hbox{{\rm I}\kern-.2em\hbox{\rm F}}}

\def\IG{\ \hbox{{\rm I}\kern-.6em\hbox{\rm G}}}

\def\IP{\hbox{{\rm I}\kern-.2em\hbox{\rm P}}}

\def\IN{\hbox{{\rm I}\kern-.2em\hbox{\rm N}}}

\def\IL{\hbox{{\rm I}\kern-.2em\hbox{\rm L}}}

\def\IM{\hbox{{\rm I}\kern-.1em\hbox{\rm M}}}

\def\smIO{\ \hbox{{\footnotesize\rm I}\kern-.4em\hbox{\footnotesize\bf O}}}
\def\smIQ{\ \hbox{{\footnotesize\rm I}\kern-.5em\hbox{\footnotesize\bf Q}}}

\def\sed{\hbox{{\sf S}\kern-.4em\hbox{\sf S}}}

\def\zed{\hbox{{\sf Z}\kern-.4em\hbox{\sf Z}}}

\def\fo{\hbox{{1}\kern-.25em\hbox{l}}}


\def\raisenot{\raise .5mm\hbox{/}}
\def\notpa{\hbox{{$\partial$}\kern-.54em\hbox{\raisenot}}}
\def\notp{\ \hbox{{$p$}\kern-.43em\hbox{/}}}
\def\notq{\ \hbox{{$q$}\kern-.47em\hbox{/}}}

\overfullrule=0pt
\tolerance=5000
\overfullrule=0pt
\twelvepoint

\twelvepoint
\date{CTP-TAMU-45/92}
\titlepage
\title{CHERN-SIMONS FORMS, MICKELSSON-FADDEEV ALGEBRAS AND
THE P-BRANES}
\vglue-.25in
\author{J. A. Dixon and M. J. Duff\foot{Research
supported in part by NSF Grant PHY-9106593.}}
\medskip
\address{Center for Theoretical Physics,
\break Texas A\&M University,
\break College Station,
\break Texas 77843}
\bigskip
\centerline{May 1992}
\abstract{In string theory,
nilpotence of the BRS operator $\d$ for the string functional relates the
  Chern-Simons
term  in the gauge-invariant antisymmetric tensor field strength
to  the
  central term in the Kac-Moody algebra.  We generalize these ideas
to p-branes with odd p and find that the Kac-Moody algebra for the string
becomes  the Mickelsson-Faddeev algebra for the p-brane.
}
\endpage
\REF\dixon{J. A. Dixon, M. J. Duff and E. Sezgin,
Phys. Lett. {\bf B279} (1992) 265.}
\REF\bergs{E. Bergshoeff, F. Delduc and E. Sokatchev,
Phys. Lett. {\bf B262} (1991) 444.}
\REF\jackiw{R. Jackiw, in {\it Lectures on Current Algebra and
Its Applications}, Princeton U.P., Princeton (1972).}
\REF\mickelssona{J. Mickelsson, Lett. Math. Phys. {\bf 7} (1983) 45.}
\REF\faddeev{L. D. Faddeev, Phys. Lett. {\bf B145} (1984) 81.}
\REF\mickelssonb{J. Mickelsson,
{\it Current Algebras and Groups},
Plenum Press, New York and London (1989). }
\REF\pressegal{A. Pressley and G. Segal, {\it Loop Groups}, Clarendon Press,
Oxford (1986).}
 \REF\siegel{W. Siegel, {\it Introduction to String Field Theory},
World Scientific, Singapore and  London (1988).}
 \REF\thorne{C. B. Thorne, Phys. Rep. {\bf 175} (1989) 1.}
\REF\duff{M. J. Duff, Class. Quant. Grav. {\bf 5} (1988) 189.}
\REF\strominger{A. Strominger, Nucl. Phys. {\bf B343} (1990) 167.}
\REF\dufflu{M. J. Duff and J. X. Lu,
Phys. Rev. Lett. {\bf 66} (1991) 1402;
Nucl. Phys. {\bf B357} (1991) 534.}
\REF\harvey{J. Harvey and A. Strominger, Univ. of Chicago preprint EFI-91-
30.}

\chapter{Introduction}

In a recent paper [\dixon],  the coupling of Yang-Mills fields to
the heterotic string in bosonic formulation was generalized to extended objects
of higher dimension (p-branes).  In particular, it was noted that for odd p the
Bianchi identities obeyed by the field strengths of the (p+1)-forms receive
Chern-Simons corrections.   In the case of the string (p=1), there is an
equality
 between the coefficient $n$ of the
Chern-Simons term $I_3(A)$ in the  antisymmetric tensor field strength
$H_3= dB_2 + n I_3(A)$,
 and the central charge $n$ of the Kac-Moody algebra
obeyed by certain operators $T^a(\sigma)$ that appear in the
gauge BRS transformations of the string functional [\bergs].
 The purpose of the present paper is to show that for
3-branes the  coefficient  of
the Chern-Simons term is equal to the coefficient  of an
Abelian extension of a $T^a(\sigma^j)$ algebra   involving new  generators
$T^a_i(\sigma^j ),\, i,j= 1,2,3$.
 The corresponding
algebras have already appeared before in the context of anomalies
[\jackiw,\mickelssona,\faddeev,\mickelssonb]
and are known in the mathematical literature as
loop algebras with a Mickelsson-Faddeev extension [\pressegal].  There
is a straightforward
generalization to $p >  3$ branes.

 In string theory, the integer $n$ also appears as a coefficient of the
Wess-Zumino-Witten term in the action, and the operators
$T^a$ can be constructed from the action [\bergs],
which is invariant under simultaneous gauge variations of the
background fields and the group coordinates.   While this
action is known for the p-branes[\dixon], the operators $T^a$ have
not yet been constructed and examined.
A second  way to get the relation is to insist on the
nilpotence of the gauge  BRS transformations of the
string field $\F$ and background fields $A$ etc.
It is this
second  method  which will here be  generalized   to the
3-brane.

\chapter{Loop Space Algebras}
In manifestly supersymmetric and $\k$-symmetric
form the heterotic string can be formulated
as a mapping from two dimensions to a target space parametrized
by variables
$X^{\m}, \q^{\a}$ and $y^m$.  We ignore $\q$ from now on.
$y^m$ are bosons parametrizing the group space.
We take the $\s$-model point of view that
there are also background fields present representing the
massless bosonic excitations of the
string.  Consider the following BRS transformation:
$$\eqalign{\d = \d_1 + \d_B} \eqn\aaab $$
Here $\d_1$ is defined by:
$$ \d_1    =
 \prod_{\m, m, \s'} \int d y^m(\s') d X^{\m}(\s')
 \Bigl \{
 \Bigl (
\int d \s
 [- \w^a T^a(\s) + \L_{\m} {d X^{\m}\over
d \s}  ] \F
 \Bigr )
 {\d  \over
\d   \F } \Big \} \eqn\del$$
where the `doubly functional' derivative is defined by:
$$ {\d \over \d \F(X)} \F(X') = \prod_{\s} \d^D[X(\s) - X'(\s)]
\eqn\fhgjd $$
and hence:
$$ \d_1 \F   =
  \int d \s
 [ - \w^a T^a(\s) + \L_{\m} {d X^{\m}\over
d \s}  ] \F    \eqn\bbbb $$
In the above, $\d_1$ is a BRS transformation which acts on
functionals of the string field $\F$,
which is itself a functional  of the string variables $X^{\m}(\s)$ and
$y^m(\s)$.  $\F$ is a string field, but we will ignore
the problems of closed string field theory here
(for reviews see e.g.  [\siegel] [\thorne] )
--in particular we ignore the dependence of $\F$ on the
reparametrization ghost fields.  The exterior derivative $d$ and
the BRS operator $\d$
are taken  to be anticommuting in this paper.
Our aim is to consider
just the Yang-Mills part of the BRS transformations of the background
fields  and the corresponding transformation of the string field.

 The variable $\s$ is the spacelike
variable on the string world sheet.
The operator $T^a(\s)$ is assumed here to depend only on
$y^m(\s)$ and functional derivatives with respect to $y^m(\s)$.
 An example of $T^a(\s)$,
for the case of the string,
can be found in [\bergs].
We shall alternate between component and form notation, for
example setting $d X^{\m} \L_{\m}  = \L_1$ etc.
 The part $\d_1$ is not separately nilpotent.
    The part $\d_B$ is separately nilpotent $(\d_B^2 =0)$
and it acts only on
the background fields
$A_{\m}^a(x)$ etc.
These BRS transformations of the background fields are:
$$ \d_B   = \int d^D x \big \{
D_{\m}^{ab} \w^b {\d \over \d A_{\m}^a}
 - {1\over2}
f^{abc} \w^b \w^c {\d \over \d \w^a}
 + [- n  A_{[\m}^a \pa_{\n]}\w^a
+ \pa_{[\m} \L_{\n]}] {\d \over \d B_{\m \n} }
$$
$$
+  [n \w^a \pa_{\m} \w^a - \pa_{\m} B_0] {\d \over \d \L_{\m } }
+ {1 \over 6}
 n f^{abc} \w^a \w^b \w^c {\d \over \d B_0 }  \big \}
\eqn\ddddd $$
Here $\L_{\m}$ is a ghost for the antisymmetric tensor field $B_{\m \n}$
and $B_0$ is a `ghost for ghost' for the ghost $\L_{\m}$.
The field $\w^a$ is the Yang-Mills Faddeev-Popov ghost.
In terms of fields this becomes, for example:
$$\d A_{\m}^a  = D^{ab}_{\m} \w^b =
   \pa_{\m}  \w^a + f^{abc} A_{\m}^b \w^c  \eqn\eeeeee $$
Alternatively we can use the notation:
$$\d A^a  = -  d \w^a - f^{abc} A^b \w^c  \eqn\ggggg $$
$$ \d \L =  n I^2_1  + d B_0  \eqn\ffffff $$
$$ \d B_2=  n I^1_2  + d \L \eqn\hhhhh $$
$$\cdots $$
  In the above the terms $I^i_{p +2 - i}$ are
the terms  of ghost number $i$ that appear in the
descent equations for the Yang-Mills fields.
In our conventions the curvature two-form is:
$$F^a  =  dA^a + {1\over2} f^{abc} A^b A^c  \eqn\iiiii $$
and it transforms as:
$$\d F^a  =   f^{abc} F^b \w^c  \eqn\jjjj $$
 The descent  equations take the form:
$$\d I^i_{p +2 - i} = d  I^{i+1}_{p +1 - i}\eqn\kk $$
so that
$$  I^0_4 =   F^a F^a  = d I^0_3 \eqn\ll $$
$$  I^0_3 =   A^a d A^a + {1\over 3} f^{abc} A^a  A^b  A^c \eqn\mm $$
$$  I^1_2 =   - A^a d \w^a \eqn\nn $$
$$  I^2_1 =   \w^a d \w^a \eqn\pp $$
$$  I^3_0 =   {1\over 6}   f^{abc}   \w^a  \w^b  \w^c \eqn\qq $$
Nilpotency follows easily using these. For example:
$$ \d^2 B_2 = n \d I^1_2 - d \d \L = 0 \eqn\fhfhfpp $$
We note that:
$$  H^0_3 = d B_2 +  n I^0_3 \eqn\rr $$
is gauge invariant:
$$ \d  H^0_3 =\d_B  H^0_3 = 0 \eqn\ss $$
$$  d  H^0_3 = I^0_4    \eqn\tt $$
We assume that the background fields and
their ghosts depend on $X$
but not on $y$, so that the action of $T$ on the
background fields and ghosts is trivial here.
We also assume that the action of $\d_B$ on the operators $T$ is
trivial, since they do not depend on the background fields.
We further assume that the string field $\F$ does not depend
on the background fields.  Note that these
operators have been defined so that $\d_1$ acts only on $\F$ and
$\d_B$ acts only on the background fields.  For example:
$$ \d_B \F    =  \d_1 A_{\m}^a    =   \d_1 \L_{\m   }     =  0 \eqn\uu $$
Calculation shows that nilpotency $(\d^2 =0)$ of $\d$
implies that the Kac-Moody algebra of the  generators $T$ has
a central term with coefficient $n$:
$$\eqalign{ [T^a(\s), T^b(\s')] = f^{abc} T^c(\s) \d(\s - \s')
+ 2 n \d^{ab} {d \over d \s} \d(\s - \s')} \eqn\kac$$

Now we want to generalize this string  case to
p-branes for   odd $p$.  The way that
    $\d^2 \F= 0 $ works  is that the
variation  $\d \L_{\m} =  n \w^a \pa_{\m} \w^a $
is compensated by the  central term in the commutator
\kac.  For higher p-branes the variation
 $\d \L_{p} =   I^2_{p}$ always involves the field
$A_{\m}^a$ in addition to the ghosts $\w^a$.  Hence
the analogue of \del \ for p-branes must have an
explicit dependence on $A^a_{\m}$
as well as  $\w^a$ and $\L_{\m_1 \cdots \m_p}$.

For example, for the 3-brane, we can accomplish
this by writing:
$$\eqalign{\d = \d_3 + \d_B}\eqn\vv $$
where $\d_3$ acts on the 3-brane wave function $\F$
$$ \d_3    =
  \prod_{\m, m, \s'^j} \int d y^m(\s') d X^{\m}(\s')
\Bigl \{
\Bigl (
  \int d^3 \s
 \{ - \w^a T^a(\s) $$
$$ - n \e^{ijk}
d^{abc} \pa_{\m} \w^a A^b_{\n} \P^{\m\n}_{ij} T^{ c}_{ k}(\s)
+ \L_{\m \n \l}
  \P^{\m\n \l}
 \} \F
\Bigr )
  {\d  \over
d   \F }  \Bigr \} \eqn\xx $$
Here we use the notation:
$$
\P^{\m\n}_{ij} =
{\pa X^{\m }\over \pa \s^{i }}
{\pa X^{\n}\over \pa \s^{j}}
\eqn\yy $$
$$
\P^{\m\n \l}  = \e^{ijk}
{\pa X^{\m }\over \pa \s^{i }}
{\pa X^{\n}\over \pa \s^{j}}
 {\pa X^{\l}\over \pa \s^{k}}
\eqn\zz $$
In the foregoing,  $\d_3$ is a BRS transformation which acts on
$\F$, which is a functional of the  3-brane  variables $X^{\m}(\s)$ and
$y^m(\s)$.    All the $T$ operators are again assumed to
involve only functions of $y^m(\s)$ and ${\d\over \d y^m(\s) }$
and hence the operators $T$ commute with $\d_B$.
The background transformations are now:
$$ \d_B   = \int d^D x \big \{
D_{\m}^{ab} \w^b {\d \over \d A_{\m}^a}
$$
$$
 - {1\over2}
f^{abc} \w^b \w^c {\d \over \d \w^a}
 + [ n I^1_4(A, \w) + d \L]_{\m \n \l \r} {\d \over \d B_{\m \n \l \r} }
$$
$$
+  [n I^2_3(A, \w) + d B_2 ]_{\m \n \l}  {\d \over \d \L_{\m \n \l} }
 + \cdots  + n I^5_0(\w)  {\d \over \d B_0 }  \big \}
\eqn\bzero$$
where
$$  I^0_5 =   d^{abc} A^a d A^b d A^c
 + \cdots \eqn\bbb $$
$$  I^1_4 =  - d^{abc} d \w^a   A^b d A^c
+ {1 \over 4} d^{abc} d \w^a   A^b f^{cde}  A^d A^e
\eqn\accc
$$
$$  I^2_3 =    d^{abc} d\w^a   A^b  d \w^c \eqn\ddd $$
$$  I^3_2 =  - d^{abc} d \w^a  d \w^b   \w^c\eqn\eee $$
 $$  I^4_1 = -  {1 \over 4} d^{abc} f^{cde}    d \w^a  \w^b  \w^d \w^e
\eqn\fff $$
$$  I^5_0 =  - {1 \over 40} d^{abc} f^{bde} f^{cfg}   \w^a  \w^d  \w^e
\w^f  \w^g \eqn\ggg $$
In particular:
$$ \d \L_{\m \n \l} =
-d^{abc} \pa_{\m} \w^a A^b_{ \n} \pa_{ \l}\w^c + \cdots
\eqn\fhdjfhj $$
By calculation, one can show that the above $\d$ is   nilpotent  if
$T^a$ and $T^a_i$
satisfy the Mickelsson-Faddeev algebra:
$$  [T^a(\s), T^b(\s')] = f^{abc} T^c(\s) \d^3(\s - \s')
- 2 n  d^{abc} \e^{i j k}
\pa_{i }  \d^3(\s - \s')
\pa'_{j} T^{c}_{ k}(\s')
 \eqn\newalga $$
$$  [T^a(\s), T^b_i(\s')] = f^{abc} T^c_i(\s) \d^3(\s - \s')
 + \d^{ab}
 \pa'_{i}  \d^3(\s - \s') \eqn\newalgb$$
$$\eqalign{ [T^a_i(\s), T^b_j(\s')] = 0 }\eqn\newalgc$$
One may verify that the Jacobi identities are satisfied by this algebra.
Note the new kind of generator $T^a_i$,
which forms
a (non-invariant) Abelian subalgebra of the $T^a$ algebra.
$T^a_i$ transforms under the action of $T^a$ like a
Yang-Mills field.

The gauge invariant field strength associated with this nilpotent
$\d_B$ is:
$$H_5 = d B_4 + n I^0_5 \eqn\hklklk $$
and it satisfies:
$$ \d H_5 = \d_B H_5 = 0 \eqn\hhh $$
$$ d H_5 = I^0_6 = d^{abc} F^a F^b F^c \eqn\jjj $$

\chapter{Spacetime Algebras}

If we take the term of $\d$ that is linear in the field
$\w^a(x)$, then its algebra is also the Kac-Moody (p=1)
or Mickelsson-Faddeev (p=3) algebra (pulled back).  This works as
follows. Define $$ \d  = \int d^4 x \w^a(x) T_{\rm tot}^a(x)
+ {\rm other\,\, terms}
\eqn\kkk $$
where the other terms are those which do not have
exactly one field $\w$ in the numerator of the transformation.

Then nilpotence of $\d$  implies  that
$$
{1 \over 2}\int d^D x  \int d^D x' \w^a(x) \w^b(x') \big \{[T_{\rm
tot}^a(x),   T_{\rm tot}^b(x') ]  $$
$$
 - \d^D(x - x') f^{abc} T_{\rm tot}^c(x) ]
\big \} \F
= n \int d^p \s I^2_{p}(X(\s))_{\m_1 \ldots \m_p}
 \P^{\m_1 \ldots \m_p} \F
\eqn\mmm $$
   Using  functional derivatives to peel off the two powers of $\w$
in the above yields a `pulled back' version of the algebra,
 which, for $p \geq 3$, has an
 $A$-dependent central extension,  determined by the form of
$I^2_{p}(X(\s))_{\m_1 \ldots \m_p}$.  For $p=1$ the extension
can be chosen to be $A$-independent because $I^2_{p}$ can be chosen to
be $A$-independent.  The $A$-dependent extension for
the $p=3$ case is somewhat
reminiscent of the situation in
four-dimensional Yang-Mills field theory with fermions
[\faddeev].  Explicitly for the 3-brane case we have:
$$[ T_{\rm tot}^a(x),  T_{\rm tot}^b(x')] \F
= \Bigl \{
f^{abc} T_{\rm tot}^c(x') \d^D(x - x')
$$
$$
+ 2 n \big [ \int d^3 \s \d^D[x - X(\s)]
d^{abc} \P^{\m \n \l}  \pa_{\m} A^c_{\n}\big ]
\pa_{\l} \d^D(x - x') \Bigr \} \F
\eqn\tot $$

\chapter{Conclusion}

Our motivation for this work was
to see how the loop space algebra of the
heterotic string can be generalized to the p-branes.
One constructs a BRS transformation that transforms
the background fields and the p-brane  functional, and
then demands that it be nilpotent.

For the string, this nilpotence  relates the
coefficient $n$ of the central extension of the Kac-Moody algebra of the
operators $T^a$ formed from the group coordinates to the
coefficient
 $n$ in   the gauge invariant field strength $$ H_{3} = d B_2 + n I_3
\eqn\nnn $$
 of the background Yang-Mills fields.

We have shown that   for the 3-brane, it is necessary
 to introduce   operators $T^a_i(\s)$  and $T^a(\s)$
which are formed from the group coordinates.
  These  operators  obey the well-known
Mickelsson-Faddeev algebra
 familiar from anomaly analysis in  four-dimensional
theories with chiral fermions. In particular the operators
$T^a_i(\s)$  transform like Yang-Mills fields  under
the action of $T^a(\s)$.  We believe that the operators $T$ obtained by
an analysis along the lines of [\bergs]
of the action in [\dixon] should provide
a realization of the Mickelsson-Faddeev algebra discussed here.
Nilpotence of the BRS transformation of the 3-brane
functional $\F$ relates the coefficient $n$ of the
(non-invariant)  Abelian  extension of the
algebra \newalga \  to the parameter $n$ in
  the gauge invariant field strength
$$ H_{5} = d B_4 + n I_5 \eqn\qqq $$
 of the background Yang-Mills fields.

We anticipate that
this procedure should easily generalize to higher $p$, and in particular to the
heterotic 5-brane
[\duff,\strominger,\dufflu,\harvey,\dixon] which in fact provided the original
impetus  for the present paper.

{\it Acknowledgment:} We enjoyed conversations with Ergin Sezgin.
\refout
\end
\bye